\documentclass[aps,pra,reprint,showpacs,groupedaddress]{revtex4-1}

\usepackage{hyperref}  
\usepackage{graphicx}  
\usepackage{dcolumn}   
\usepackage{bm}        
\usepackage{amssymb}   
\usepackage{verbatim}  
\usepackage{epstopdf}
\usepackage{color}

\newcommand{\eq}[1]{(\ref{#1})}
\newcommand{\beq}{\begin{equation}}
\newcommand{\eeq}{\end{equation}}
\newcommand{\half}{\hbox{$1\over2$}}

\begin{document}

\title{Nonlinearity from quantum mechanics:\\ Dynamically unstable Bose-Einstein condensate in a double-well trap}
\author{Juha Javanainen}
\affiliation{Department of Physics, University of Connecticut,
Storrs, CT 06269-3046}
\begin{abstract}
We study theoretically an atomic Bose-Einstein condensate in a double-well trap both quantum mechanically and classically under conditions such that in the classical model an unstable equilibrium dissolves into large-scale oscillations of the atoms between the potential wells. Quantum mechanics alone does not exhibit such nonlinear dynamics, but measurements of the atom numbers in the potential wells may nevertheless cause the condensate to behave essentially classically.
\end{abstract}
\maketitle

Even though quantum mechanics is supposedly the fundamental theory, a classical description succeeds for a large host of systems. When the classical model is nonlinear, apparent discrepancies with linear quantum mechanics readily arise. We have earlier~\cite{JAV08} studied stationary states of a Bose-Einstein condensate (BEC) in an optical lattice. Classical theory predicts a nonlinear soliton, while quantum mechanics does not. Of course, solitons are seen experimentally~\cite{EIE04}. Our resolution of this dilemma is that the measurements probing the soliton will bring it about~\cite{JAV08}. 

Here we study a time-dependent case in which quantum mechanics and classical mechanics seem to disagree, namely, dynamical instability. Classically, the signature of an instability is that a small deviation from an unstable equilibrium grows exponentially in time. Quantum mechanically, under unitary time evolution, the distance between any two states remains constant.  

A Bose-Einstein condensate in a symmetric double-well potential makes our example. A convenient quantum description exists~\cite{MIL97}, and can be easily solved numerically for a large number of atoms~\cite{JAV99}. There is a corresponding classical model, too, exactly as in quantized and classical descriptions of an electromagnetic field. It is nonlinear, and has been discussed in particular as an example of population trapping~\cite{MIL97,RAG99}, asymmetric oscillations of the atoms between the two wells. Population trapping has also been seen experimentally~\cite{ALB06}. The classical system may have an unstable equilibrium, and a smallest deviation from the equilibrium can lead to large-scale oscillations of the atoms between the two sides of the potential well~\cite{SHR08}. The question is, what does quantum mechanics have to say about such oscillations.

We describe a preparation that leaves the classical model in an unstable state. The oscillations of the atoms differentiate between the two potential wells. Since quantum mechanics strictly preserves the symmetry between the sides of the trap, there seemingly cannot be any such oscillations. But the broken symmetry in the classical case~\cite{JAV96,RUO97,KAN05,JAV08} once more~\cite{JAV96,RUO97,JAV08} correctly suggests that including a model for the observations of the atoms will cure the discrepancy. We do this much as before~\cite{RUO97}, and solve the theory using quantum trajectory simulations~\cite{TIA92,DAL92,DUM92,GAR99}. The quantum system, thus amended, will not only  display the classical behavior, but, in analogy to Ref.~\cite{JAV08}, the measurements are seen to literally cause the classical dynamics.

We study a double-well trap for bosonic atoms within the conventional two-mode approximation~\cite{MIL97}.\ The Hamiltonian is
\vspace{-2pt}
\beq
\frac{H}{\hbar} = -J (b^\dagger_r b_l + b^\dagger_l b_r) + 
U (b^\dagger_r b^\dagger_rb_r b_r + b^\dagger_l b^\dagger_l b_l b_l )\,.
\label{HAM}
\vspace{-2pt}
\eeq
Here $b_l$ and $b_r$ are the annihilation operators for the atoms in the ``left'' and ``right'' wells, $J>0$ is the amplitude for tunneling between the wells, and $U$ characterizes the strength of the atom-atom interactions. The total number the of atoms 
$
\hat{N} =b^\dagger_r b_r + b^\dagger_l  b_l
\label{NCD}
$
 is a constant of the motion, and its value is denoted by $N$.

The corresponding classical field theory is obtained by taking the Heisenberg equations of motion for the boson operators, e.g., $i\dot{b}_r = -J b_l+ 2 U b^\dagger_r b^\dagger_r b_r$,  and declaring that in the equations of motion the operators are $c$-numbers, no longer operators. We then have, for instance, $i\dot{b}_r = -J b_l+2 U |b_r|^2 b_r$. 
This classical system has only two relevant dynamical variables~\cite{RAG99} that we pick as
\vspace{-2pt}
\beq
z = (|b_r|^2-|b_l|^2)/N,\quad \varphi = \arg (b_rb^\dagger_l)\,,
\label{ZANG}
\eeq 
relative population imbalance and phase difference of the condensates between the right and left traps. The variables $z$ and $\varphi$ make a canonical pair, and their evolution is governed by the classical Hamiltonian
\vspace{-2pt}
 \beq
{\cal H} = -2J\sqrt{1-z^2}\,\cos\varphi +\chi (1+z^2);\, \chi=NU\,.
 \label{CLH}
 \eeq
\vspace{-10pt}
 
The variables $z$ and $\varphi$ scale away the dependence on total atom number $N$, but $N$ is still is part of the classical Hamiltonian~\eq{CLH} in that $\chi\propto N$. If ever, the classical approximation can be accurate only for a large number of atoms, when the discreteness of the atoms may be ignored. In the  classical limit  of the quantum problem embodied in Eq.~\eq{HAM} one should simultaneously take $N\gg1$, and scale  the atom-atom interaction strength $U$ so that $\chi=NU$ remains constant. As the interaction strength $U$ can be adjusted using the Feshbach resonance~\cite{FBRES} and the atom number can also be varied experimentally, the two-well model represents a realistic system that can be tuned between the quantum and classical limits.

\begin{figure}
 \begin{center}
\includegraphics[width=8.5cm]{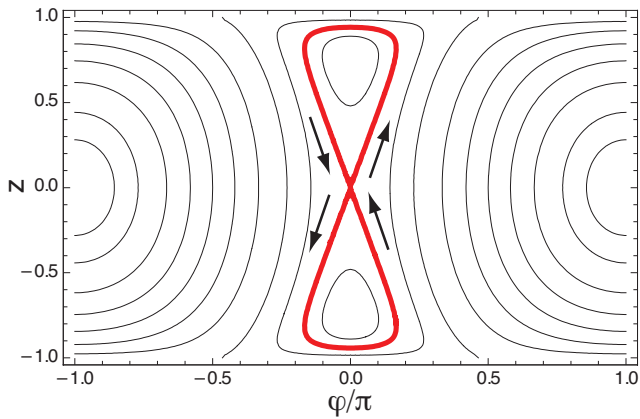}
\end{center}
\vspace{-20pt}
\caption{(Color online) Phase space portrait for the classical model of a two-well system for the parameter values $\chi/J = -1.5$. The thick line marks the homoclinic orbits; the arrows indicate the direction of flow in phase space. }
\label{PHP}    
\vspace{-15pt}
\end{figure}

The classical Hamiltonian~\eq{CLH} has two stationary states in the region $(z,\varphi)\in[-1,1]\otimes[0,2\pi)$, namely, $(z,\varphi) = (0,0)$ and $(0,\pi)$. Both are dynamically stable for $|\chi|/J<1$; for $\chi/J>1$ the steady state $(z,\varphi)=(0,\pi)$ is unstable, while for $\chi/J<-1$ the unstable state is $(z,\varphi)=(0,0)$. Figure~\ref{PHP} shows a phase-space portrait, contours of constant ${\cal H}(z,\varphi)$. Since the Hamiltonian is a constant of the motion, the phase-space coordinate $(z,\varphi)$ is constrained to move along a constant-energy contour. Figure~\ref{PHP} is for the case with $\chi/J=-1.5$, whereupon the point $(z,\varphi) = (0,0)$ is a dynamically unstable equilibrium, a hyperbolic fixed point. There are two directions in phase space in which the system first recedes exponentially in time from the point $(0,0)$ and then returns to the fixed point, completing a loop called homoclinic orbit. 

To facilitate an unbiased comparison between classical and quantum dynamics, we envisage an experimentally feasible preparation that works in the same way for both cases.  Thus, the system is first prepared to the lowest-energy state in the presence of repulsive atom-atom interactions, $\chi/J>1$, and at the time $t=0$ the sign of the atom-atom interaction is suddenly flipped.

Classically, the lowest-energy state for positive interactions is $(z,\varphi)=(0,0)$, with half of the atoms on each side of the trap and the same phase for the condensates on both sides; for $\chi/J <-1$ the same state is the unstable equilibrium. Whether in numerical or physical experiments, a classical system will not stay in a dynamically unstable state, but some form of noise will invariably launch the instability. In the example of Fig.~\ref{PHP} even a minute amount of noise can cause three distinct macroscopic behaviors~\cite{SHR08}. If the system starts inside the homoclinic orbit with $z>0$, in the absence of further noise it will stay inside the same orbit. Correspondingly, the atoms stay predominantly in the right trap. The same situation but favoring the left trap occurs if the system starts inside the homoclinic orbit with $z<0$. The third alternative is that the system starts (slightly) outside of the homoclinic orbits. Then symmetric oscillations of the atoms from side to side will result.

In the quantum case, the preparation puts the system in the ground state for repulsive atom-atom interactions. This state interpolates from the situation with all atoms in the single-particle ground state of the double-well trap, $|\psi\rangle \propto (b^\dagger_r+b^\dagger_l)^N|0\rangle$, to the number state with half of the atoms in each side, $|\psi\rangle\propto {b_r^\dagger}^{N/2}{b_l^\dagger}^{N/2}|0\rangle$, as the interaction strength ranges from $U=0$ to $N U/J\rightarrow\infty$~\cite{MIL97,JAV99}. It is not a stationary state after the sign of the interaction is flipped. However, since both the ideal initial state and the Hamiltonian are invariant under the exchange of the site labels $l$ and $r$, Hamiltonian time evolution alone cannot produce any overt difference between the two sides of the trap either: every quantum expectation value remains unchanged under the exchange of  $l$ and $r$. Moreover,  and  in contrast to the classical case, a small error in the initial state will not cause runaway oscillations of the atoms between the two sides. To find a counterpart of classical instability we look elsewhere, into measurements.

In our thought experiments we assume in close analogy with Refs.~\cite{RUO97} and \cite{RUO98} that, perhaps by utilizing far-off resonant light scattering, a situation has  been set up whereby two photon detectors, ``left'' and ``right'', give photon counts as a result of the presence of the atoms in the left and right traps at the rates $R_{r} = \Gamma \langle{b^\dagger_r b_r}\rangle$ and  $R_{l} = \Gamma \langle{b^\dagger_l b_l}\rangle$. The constant $\Gamma$ could depend on geometry of the experiment and the tuning and intensity of the probe light. Here we make a technical assumption that removes various secondary complications from the picture: no photons are missed, but there is instead a one-to-one correspondence between photon detections and their back-action on the condensate. Formally, we model the photon detections by the operators
\vspace{-5pt}
\beq
L_{r,l}=\sqrt{\Gamma/2}\, \sqrt{b^\dagger_{r,l}b_{r,l}}
\vspace{-5pt}
\eeq
such that the Liouville-von Neumann equation of motion of the density operator of the BEC, $\rho$, is amended by an addition in the usual Lindblad form~\cite{GAR99},
\beq
\vspace{-5pt}
\dot\rho =- \frac{i}{\hbar}[H,\rho]+\sum_{i=l,r} (2 L_i\rho L^\dagger_i - L^\dagger_i L_i\rho - 
\rho L^\dagger_i L_i)\,.
\label{MODSCH}
\eeq

The equation of motion~\eq{MODSCH} is unraveled using standard quantum trajectory simulations~\cite{TIA92,DAL92,DUM92,GAR99}. For reasons of efficiency and accuracy the practical algorithm is more complicated~\cite{DUM92,GAR99} than our conceptual discussion~\cite{DAL92}, but the idea is as follows. Suppose we have a state vector for the system at time $t$, $|\psi(t)\rangle$. To obtain the state vector a short time $dt$ later, we first evolve the state according to
\vspace{-5pt}
\beq
\frac{d}{dt} |\psi\rangle = \left(-\frac{i}{\hbar} H - \sum_i L^\dagger_i L_i\right)|\psi\rangle\,.
\eeq
to obtain $|\tilde\psi(t+dt)\rangle$.
In our particular example the combination of the Lindblad operators give $\sum_i  L^\dagger_i L_i = \half\Gamma \hat{N}$, so that the effect of the term added to the Schr\"odinger equation is simply to damp the state vector by the factor $e^{-\half N\Gamma\,dt}$. At the end of the integration over $dt$, the probability for a photon count on the right detector is calculated, $dP_r = dt\,R_r = dt\, \Gamma \langle{b^\dagger_r b_r}\rangle$, and $dP_l$ analogously. Using a random number generator, the algorithm decides whether a photon count happens on either detector. If the decision is that the right detector $r$ clicked, the state vector at time $t+dt$ will be $|\psi(t+dt)\rangle = L_r |\psi(t)\rangle \propto \sqrt{b^\dagger_r b_r}\,|\psi(t)\rangle$; if the detector $l$ reported, the result is $|\psi(t+dt)\rangle \propto 
\sqrt{b^\dagger_l b_l}\,|\psi(t)\rangle$; if neither detector reported, the state at $t+dt$ is $|\psi(t+dt)\rangle=|\tilde\psi(t+dt)\rangle$; and for a short enough time steps $dt$ the possibility that both detectors clicked is negligible. Finally, the state $|\psi(t+dt)\rangle$ is normalized, and the next time step commences.

A collection of stochastic state vectors obtained in this way could be used to compute expectation values over the density operator $\rho$~\cite{TIA92,DAL92,DUM92,GAR99}. More to the present point, it can be argued that the sequence of photon counts produced along with an individual realization of a quantum trajectory $|\psi(t)\rangle$ is a representative example of a sequence of photon counts that one would see in one run of a conforming experiment. This idea permeates the literature on quantum trajectory simulations~\cite{GAR99}, but for completeness we enunciate the underlying (meta)physical assumption~\cite{JAV86}. Namely, by construction of the quantum trajectory algorithm  the multitime correlation functions for the photon counts from repeated runs of the simulation would be the same as the correlation functions from repeated real experiments as predicted from quantum mechanics under the Markov approximation and  the quantum regression theorem~\cite{GAR99}. If one reasonably assumes that it is not possible to tell the difference between the realizations of two stochastic processes with the same correlation functions, quantum trajectory simulations and experiments should produce indistinguishable sequences of photon counts.

We may now consider what would happen in an individual experiment in which the quantum system starts from the state whose classical counterpart is dynamically unstable. If we were to analyze an actual experiment, in the present setup the only information available would be the timing of the photon counts. We would have to reconstruct the counting rates at both detectors from the observed photon counts, and infer from those the numbers of the atoms in each potential well. In a simulation we have the shortcut that the instantaneous photon counting rates $R_r$ and $R_l$ are on hand, and we may use them to find  a representation of the population imbalance. Analogously to Eqs.~\eq{ZANG} we compute
\beq
z(t) ={\langle\psi(t)|( b^\dagger_r b_r - b^\dagger_l b_l) |\psi(t)\rangle}/{N}\,.
\label{EMPZ}
\eeq
In the simulations we also define the phase difference between the condensates by closely mimicking Eq.~\eq{ZANG},
\beq
\varphi(t) = \arg\langle \psi(t)|b_r b^\dagger_l|\psi(t)\rangle\,.
\label{EMPF}
\eeq
It is possible to measure experimentally the phase difference between the condensates~\cite{SAB05}, but in our scheme $\varphi(t)$ from Eq.~\eq{EMPF} is an auxiliary quantity with no direct operational meaning. 

\begin{figure}
 \begin{center}
\hspace{-0.5cm}
\includegraphics[width=8.5cm]{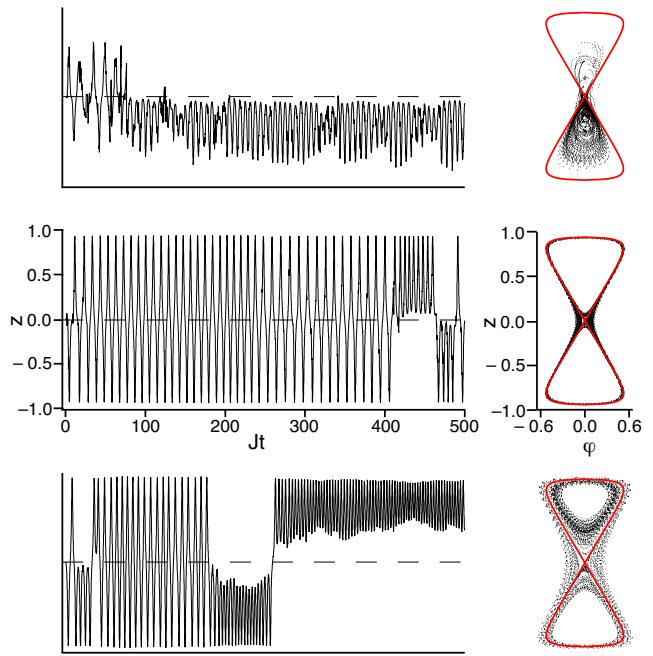}
\end{center}
\vspace{-10pt}
\caption{(Color online) Time dependence of population imbalance $z(t)$ and the phase space trajectory $z(\varphi)$ from three quantum trajectory simulations that differ in the frequency of the observations by photon counting; from top to bottom, $N\Gamma/J = 1$, $100$, and $10^4$. The trajectory  in phase space is plotted as discrete points, 5000 of them; the solid line shows the same homoclinic orbits as Fig.\ref{PHP} .}
\label{SUMRY}    
\vspace{-15pt}
\end{figure}

Example results are given in Fig.~\ref{SUMRY}. Here we have the atom number $N=10^4$ and the interaction strength $\chi/J = NU/J = -1.5$.  Each data set shows the outcome of one quantum trajectory simulation. The difference is in the frequency of detection of the atom numbers, set for instance by varying the intensity of the detection light; from top to bottom we have $N\Gamma/J=1$, $100$, and $10000$. The graphs on the left present the population imbalance as a function of time computed from Eq.~\eq{EMPZ},  while the graphs on the right show the population imbalance $z(t)$ versus the angle $\varphi(t)$, Eq.~\eq{EMPF}, as points for all of the 5000 time samples that were also used to plot $z(t)$.

First look at the middle data set with $N\Gamma/J=100$. In the plot of $z(t)$ all three behaviors anticipated for the classical system are apparent at once: symmetric oscillations between the wells, and oscillations in which a majority of the atoms stays in one or the other of the wells.   A classical system starting from a random initial condition and thereafter evolving deterministically would stick with one type of behavior. We attribute the switching to quantum noise and/or noise due to the back-action of the measurements. On the other hand, the plot of $z(\varphi)$ is as if the system stayed close to the homoclinic orbits, something that the classical system is expected to do.

The top data set differs in the frequency of observations, $N\Gamma/J=1$. We qualitatively assign a period of the oscillations in population trapping, approximately half of the period of those oscillations in which the atoms go back and forth between the traps, as the characteristic time scale for the oscillations. Then the top panel corresponds to about five photon counts per characteristic time, the middle panel to about five hundred. There is still a qualitative resemblance to the classical dynamics  in the results for $N\Gamma/J=1$, but the quantitative agreement is quite poor compared to the $N\Gamma/J=100$ case. When the rate of observations is further decreased, the resemblance to classical dynamics deteriorates further. In a very real sense the observations cause the classical behavior: for the classical dynamics to emerge, the state has to be observed frequently enough that the classical behavior can be resolved.

In the bottom data set we have $N\Gamma/J=10^4$, which means that several photons are recorded per each atom during a characteristic time of the oscillations of the atoms between the traps. Although the noise is enhanced, the classical behavior is still clearly discernible. Our interpretation is that, while increasingly aggressive measurements will eventually destroy the classical behavior, the system is highly resilient against the noise from the back-action of the measurements.

Our particular model for continuous measurements of the atom number was picked for numerical expediency, as quantum trajectory simulations boil down to solving the Schr\"odinger equation. Real experiments on the instability would not necessarily comply. For instance, suppose one has inside a Mach-Zehnder interferometer light that is far-off resonance from any atomic transition, and that the interferometer is initially balanced so that all light comes out from one port. When one inserts an atomic sample inside one arm of the interferometer, the refractive index from the atoms make light to come out of the formerly dark port. But the amplitude of this light is proportional to the number of the atoms, so that the intensity, and photon counting rate, is proportional to the square of the number of the atoms not atom number~\cite{RUO98}.

The notion that measurements or other environmental influences  may cause a quantum system behave approximately classically is by no means new~\cite{ZUR03}, and quantum trajectory type methods have been used to study the emergence of classical chaos in quantum systems~\cite{SPI94,BRU97,GHO03,EVE09}. The unusual feature about our dynamically unstable model is that a dramatic macroscopic behavior is triggered entirely by measurement back-action. We do not enter into details of different measurement schemes, for one thing because we believe that within reason (many atoms, measurements not too weak or strong) classical physics always emerges~\cite{BRU97,GHO03,EVE09}. However, although the smaller atom numbers dictated by the constraints of the numerics make the results less striking, we have verified the same qualitative behaviors as in our explicit examples also in the case when the photon counting rates are proportional to the squares of the atom numbers.

Making the sides of the trap and the measurement scheme symmetric enough that quantum phenomena rather than technical imperfections dictate the behavior of the atoms may be a challenge in real experiments. On the other hand, the assumption that no scattered photon is lost without a detection event is clearly unimportant for the basic principle. For instance, the detection of the classical behavior suffers because of the less-than-unit efficiency of a photon counter, but information about the atoms numbers gets conveyed to the detectors and there is a back-action from the measurements. Analysis of the effects of nonidealities of the system and of the measurements is an interesting problem area, but we leave it for future work.

We have studied the behavior of a Bose-Einstein condensate in a double-well trap under the conditions when the classical model of the system has an unstable steady state and exhibits large-scale nonlinear dynamics as a result. A priori, the corresponding quantum system  does no such thing, but when the state is monitored the classical nonlinear dynamics nonetheless emerges. That a quantum mechanical calculation gives you nothing without a description (at least an implied description) of the measurements may be trivial in itself. We think, though, that we also have a practical point: When one deals with a system that is intermediate between quantum and classical mechanics, a careful description of the measurements is a necessary part of the analysis.

This work is supported in part by NSF (PHY-0651745). The author acknowledges a very helpful discussion with Klaus M{\o}lmer.

\end{document}